\documentclass[acmsmall]{acmart}

\usepackage{markdown}
\usepackage[disable]{todonotes}

\AtBeginDocument{%
  }

\setcopyright{cc}
\setcctype{by-sa}
\acmJournal{TORS}
\acmYear{2025} 
\acmMonth{8} 
\acmVolume{TBA}
\acmNumber{TBA}
\acmArticle{1} 
\acmPrice{}
\acmDOI{10.1145/3759261}

\makeatletter
\def\@mkbibcitation{\bgroup
  \let\@vspace\@vspace@orig
  \let\@vspacer\@vspacer@orig
  \def\@pages@word{\ifnum\getrefnumber{TotPages}=1\relax page\else pages\fi}%
  \def\footnotemark{}%
  \def\\{\unskip{} \ignorespaces}%
  \def\footnote{\ClassError{\@classname}{Please do not use footnotes
      inside a \string\title{} or \string\author{} command! Use
      \string\titlenote{} or \string\authornote{} instead!}}%
  \def\@article@string{\ifx\@acmArticle\@empty{\ }\else,
    Article~\@acmArticle\ \fi}%
  \par\medskip\small\noindent{\bfseries ACM Reference Format:}\par\nobreak
  \noindent\bgroup
    \def\\{\unskip{}, \ignorespaces}\authors\egroup. \@acmYear. \@title
  \ifx\@subtitle\@empty. \else: \@subtitle. \fi
  \if@ACM@nonacm\else
    \if@ACM@journal@bibstrip
       \textit{\@journalNameShort},
       Just Accepted (\@acmPubDate),
       \ref{TotPages}~\@pages@word.
    \else
       In \textit{\@acmBooktitle}%
       \ifx\@acmEditors\@empty\textit{.}\else
         \andify\@acmEditors\textit{, }\@acmEditors~\@editorsAbbrev.%
       \fi\
       ACM, New York, NY, USA%
         \@article@string\unskip, \ref{TotPages}~\@pages@word.
    \fi
  \fi
  \ifx\@acmDOI\@empty\else\@formatdoi{\@acmDOI}\fi
\par\egroup}
\makeatother

\begin{document}
\makeatletter
\fancypagestyleassign{origstd}{standardpagestyle}
\fancypagestyleassign{origfirst}{firstpagestyle}
\fancypagestyle{standardpagestyle}[origstd]{
    \fancyfoot[RO,LE]{\footnotesize \@journalNameShort, Just Accepted (\@acmPubDate).}
}
\fancypagestyle{firstpagestyle}[origfirst]{
    \fancyfoot[RO,LE]{\footnotesize \@journalNameShort, Just Accepted (\@acmPubDate).}
}
\makeatother

\title{Recommending With, Not For}
\subtitle{Co-Designing Recommender Systems for Social Good}

\author{Michael D. Ekstrand}
\email{ekstrand@acm.org}
\orcid{0000-0003-2467-0108}
\author{Afsaneh Razi}
\email{ar3882@drexel.edu}
\orcid{0000-0001-5829-8004}
\author{Aleksandra Sarcevic}
\email{as3653@drexel.edu}
\orcid{0000-0003-2126-8527}
\affiliation{
\institution{Drexel University}
\city{Philadelphia}
\state{PA}
\country{USA}
}
\author{Maria Soledad Pera}
\email{M.S.Pera@TUDelft.nl}
\orcid{0000-0002-2008-9204}
\affiliation{
\institution{Delft University of Technology}
\city{Delft}
\country{NL}
}
\author{Robin Burke}
\email{robin.burke@colorado.edu}
\orcid{0000-0001-5766-6434}
\affiliation{
\institution{University of Colorado}
\city{Boulder}
\state{CO}
\country{USA}
}
\author{Katherine Landau Wright}
\email{KLWright@uoregon.edu}
\affiliation{
\institution{University of Oregon}
\city{Eugene}
\state{OR}
\country{USA}
}

\renewcommand{\shortauthors}{Ekstrand et al.}

\begin{abstract}
Recommender systems are usually designed by engineers, researchers, designers, and other members of development teams. These systems are then evaluated based on goals set by the aforementioned teams and other business units of the platforms operating the recommender systems.
This design approach emphasizes the designers' vision for how the system can best serve the interests of users, providers, businesses, and other stakeholders.
Although designers may be well-informed about user needs through user experience and market research, they are still the arbiters of the system's design and evaluation, with other stakeholders' interests less emphasized in user-centered design and evaluation.
When extended to recommender systems for social good, this approach results in systems that reflect the social objectives as envisioned by the designers and evaluated as the designers understand them. Instead, social goals and operationalizations should be developed through participatory and democratic processes that are accountable to their stakeholders.
We argue that recommender systems aimed at improving social good should be designed \emph{by} and \emph{with}, not just \emph{for}, the people who will experience their benefits and harms. That is, they should be designed in collaboration with their users, creators, and other stakeholders as full co-designers, not only as user study participants.
\end{abstract}

\ccsdesc[500]{Human-centered computing~Participatory design}
\ccsdesc[500]{Human-centered computing~HCI design and evaluation methods}
\ccsdesc[500]{Information systems~Recommender systems}
\ccsdesc[300]{General and reference~Evaluation}

\keywords{co-design, co-evaluation, participatory design, recommender systems}

\maketitle

\section{Introduction}
\label{sec:intro}

Recommender systems have a substantial impact not only on the people who directly use them, but also on other stakeholders\footnote{We use the term ``stakeholder'' for familiarity and consistency with the multistakeholder recommendation literature \citep{burkeMultisidedFairnessRecommendation2017, abdollahpouriRecommenderSystemsMultistakeholder2017}. However, we note that this term is contested \citep{sharfsteinBanishingStakeholders2016} and obscures the varying power relationships that participatory design in its full form seeks to directly confront.} and on the broader information and economic ecosystems in which they are embedded.
As people use recommender systems to locate information, entertainment, goods, and services that meet their often personalized and contextual needs, the system has a pivotal role in providing or withholding exposure to providers and their content, influencing user perceptions of the information space and aiding or impeding access to different types of information. In practice, this means that recommender systems have the potential to promote diverse and equitable marketplaces \citep{mehrotraFairMarketplaceCounterfactual2018}, influence users towards personally- and socially-beneficial activity \citep{starkeEffectiveUserInterface2017,knijnenburgRecommenderSystemsSelfActualization2016}, and advance other pro-social outcomes.
However, recommender systems can also entrench established players, facilitate the spread of disinformation and propaganda, promote social division or unsustainable consumption practices, and generally catalyze societal harm~\citep{belkinEthicalPoliticalImplications1976,milanoRecommenderSystemsTheir2020}.

Recommender systems research and development is carried out by the engineers, researchers, designers, and other development team members, typically employed by the owner of the platforms where the systems will be deployed (collectively, the recommender \textit{designers}).
These designers are usually the primary agents in designing and evaluating recommendation experiences, and users are involved only as research subjects or occasionally through early-stage design activities such as focus groups.
It is rare for users to be involved in the design of the system or in the design and interpretation of the evaluations used to validate it.
Because evaluation and design are often based on observation of behavior,  users have little opportunity to influence design or evaluation outside the behavioral measurements defined by the designers.
\citet{ekstrandBehaviorismNotEnough2016} critiqued this trend, calling for increased involvement of explicit user preferences, while \citet{seaverCaptivatingAlgorithmsRecommender2018} examines recommendation and behavior through the lens of ``trapping.''
Current practice, however, overwhelmingly emphasizes the designers' notions of what it means for the recommender system and its effects to be ``good,'' both for individual users and for society.
As the recommender systems community looks beyond providing user and business benefit to projects that advance social good, the importance of incorporating diverse perspectives into research and development only increases.

Recent work has sought to qualitatively understand the various human factors involved in recommendation \cite{peraHumanFactorsUser2024} and what users \citep{harambamDesigningBetterTaking2019, smithExploringUserOpinions2020}, artists \citep{ferraroWhatFairExploring2021}, and other stakeholders want from recommendation \citep[\S4.4]{bauerEvaluationPerspectivesRecommender2024}.
\citet{strayBuildingHumanValues2024} discussed the need for eliciting values from a wide range of perspectives and disciplines and incorporating them into the recommender design process, but the field so far has few examples of how to implement this in practice.

In this paper, we examine \textbf{who is, or can be, meaningfully empowered to determine the goals, methods, and design of recommender systems}, both with respect to their ability to meet individual users' needs and their impact on other people, organizations, and society.
We argue that recommender systems should be designed in collaboration with users, providers, and other stakeholders as full co-designers instead of merely user study participants.
Our claim is that to effectively support both individual and social good through recommendation, designers must fully involve the many different stakeholders of the recommender ecosystem --- including the users it will serve, the providers whose content visibility it will affect, and others impacted by the system --- in the process of designing and evaluating recommender systems.
We build on and extend prior arguments that explicit user goals and preferences should be elicited and considered
\citep{ekstrandBehaviorismNotEnough2016, lyngsTellMeWhat2018} to propose engaging users, providers, and other stakeholders as \textbf{designers and evaluators} of recommender systems \cite{charisiArtificialIntelligenceRights2022}.
Recommender system design can and should be a collaborative, participatory effort where designers \textit{co-design} the system with the people it will affect, an approach we call \textbf{participatory recommendation}.

We develop this argument by first reviewing common practices, user-centric developments, and social concerns in recommender systems (\S\ref{sec:rs-eval}) and providing a brief primer on the principle of co-design from human-computer interaction (\S\ref{sec:codesign}).
We then lay out our vision for co-design of recommender systems (\S\ref{sec:corecsys}), and describe the impact of this vision through several case studies across diverse domains and applications (\S\ref{sec:cases}).

\section{Current Recommender System Design and Evaluation Practice}
\label{sec:rs-eval}

Current practice in recommender system design and evaluation typically starts with the designer as the source of ideas and requirements for the recommendation experience.
We briefly review this practice and the associated literature as a foundation for our argument about re-negotiating the power relationships in recommender design to enable participatory recommendation.

\subsection{Design Lifecycle and Participants}

\begin{figure*}[tb]
\includegraphics[width=\textwidth]{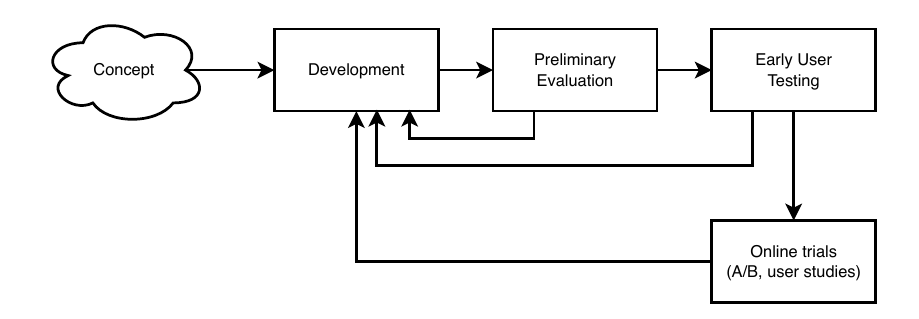}
\Description{
    A block diagram of recommender system process.
    In the top row, there is cloud on the left side labeled “Concept.”
    Arrows show the process progressing from “Concept” through “Development,” “Preliminary Evaluation,” and “Early User Testing,” all on the top row, and then to “Online trials (A/B, user studies)” in the bottom right.
    Arrows go back to Development from all three subsequent boxes.
}
\caption{Diagram of typical recommender system design and evaluation process.}
\label{fig:process}
\end{figure*}

Recommender systems evaluation researchers have presented several perspectives on recommender system development and evaluation with slightly different terms and taxonomies \citep{knijnenburgExplainingUserExperience2012,gunawardanaEvaluatingRecommenderSystems2022, zangerleEvaluatingRecommenderSystems2022}. However, the fundamental approach remains relatively consistent, with the typical lifecycle for designing and evaluating a recommender system proceeding as follows (Fig. \ref{fig:process}):

\begin{enumerate}
    \item The designers identify an opportunity for new or improved recommendations. This opportunity may emerge from their own insight and experience, from user studies identifying gaps in the user experience, or from other business concerns such as marketing, contractual obligations, and regulatory compliance.
    \item The designers use existing data to perform a preliminary offline evaluation of the idea, often iterating on several design options and configurations.
    \item The designers may involve small groups of users in early-stage product testing, particularly if they have strong human-computer interaction capabilities.
    \item The designers test the new or revised system online with users, using techniques such as behavioral A/B testing with user activity and business metrics (e.g., clicks, engagement time, or purchases) and perceptual testing with user studies incorporating surveys and explicit user feedback.
\end{enumerate}

This process is iterative, with the results of one design or testing exercise informing the next step or iteration.
The specifics of this process can differ among platforms, research teams or projects, with the level of human involvement increasing or decreasing at various stages. For example, \citet{harambamDesigningBetterTaking2019} involved users in a focus group study to understand user goals and desires for recommendation explanation capabilities.
The key point related to our argument in this paper is that the dominant approach involves users solely or primarily as \textit{evaluators} (sometimes explicitly, through user studies \citep{knijnenburgExplainingUserExperience2012}, but often implicitly through A/B trials \citep{kohaviTrustworthyOnlineControlled2020}) of designs originating from the designers. The types of feedback users can provide are typically also circumscribed by the evaluation mechanisms provided by the designers.
\citet{ekstrandBehaviorismNotEnough2016} critiqued this predominant approach, especially in its implicit, behavioral form, arguing that greater explicit user involvement across the design and evaluation lifecycle will enable more useful and pro-social recommendation technology.

A significant limitation of engaging with users only as recipients and evaluators of the designers' ideas is that it restricts the set of \textit{values} \citep{strayBuildingHumanValues2024} that are to be incorporated into the recommendation experience: the design team or platform's values take priority, supplemented with their view of other stakeholder values.
When considering recommender systems for social good, some goals may be easily discoverable, such as energy savings \citep{starkeEffectiveUserInterface2017} and sustainability \citep{tomkinsSustainabilityScaleBridging2018, patroSafetySustainabilityDesigning2020}. Still, design and evaluation need mechanisms to incorporate a much broader set of perspectives on what social good actually is and how recommendations can contribute to that good to develop and deploy systems that are beneficial for humanity and the natural, communal, and socio-technical ecosystems it inhabits.

\subsection{Multistakeholder Recommendation}

Users are not the only people with a stake in recommender systems and their behavior and impact.
\textit{Multistakeholder recommendation} \citep{abdollahpouriRecommenderSystemsMultistakeholder2017,burkeMultisidedFairnessRecommendation2017,sonboliMultisidedComplexityFairness2022} takes into account the needs and desires of a broader set of stakeholders, who directly or indirectly stand to benefit from (or be harmed by) the recommendation process. Common stakeholders include \emph{providers} (the artists, authors, and other creators of the items being recommended, and who benefit from their exposure to potentially interested users), \emph{subjects} (the people discussed by an item, such as communities affected by the events described by a news article), \emph{platforms} or \emph{systems}, \emph{developers}, and other groups.
\citet{burke:post-userist} take this further, arguing for a \textbf{relational} approach to understanding recommender systems that emphasizes the relationships between many different groups affected by the system, including their relationships to each other (e.g., the relationship between readers and authors) affected but not mediated by the system.

The last decade has seen increased attention to the concerns of various stakeholders within the traditional evaluation paradigm, such as assessing \textit{provider exposure} \citep{singhFairnessExposureRankings2018,biegaEquityAttentionAmortizing2018,diazEvaluatingStochasticRankings2020,rajMeasuringFairnessRanked2022} and examining the distributions of effects between and within different stakeholder groups \citep{ekstrandDistributionallyinformedRecommenderSystem2024}.
However, deep engagement with these groups that sees their values and goals reflected in the design and evaluation process is still in its infancy \citep[\S4.4]{bauerEvaluationPerspectivesRecommender2024}.

\subsection{Design and Evaluation for Specific Users and Contexts}

In the pursuit of adaptation and personalization, and with the advances of technology, recommender systems --- particularly those aiming to promote social good --- are now more than ever focused on the ``human'' and their values and desires, and not just on their role as a ``user.'' This expansion of human concerns is in addition to considerations inherent to particular recommendation contexts \cite{peraHumanFactorsUser2024}. This shift, however, presents new challenges and demands, expanding the scope of more traditional design and evaluation practices. Instead of solely designing generic solutions, optimizing algorithms, and evaluating in terms of average accuracy, there is now a need to consider human factors (e.g.,  personality, emotional and cognitive states) and specific context requirements that influence every stage of user and item modeling, system design, evaluation, and ultimately deployment.

Research and industry have produced a variety of design and evaluation practices that best fit the needs of specific user groups or contexts. For example, some recommendation strategies are specifically created for users with different cognitive abilities, at various developmental stages, or for use in settings such as education or healthcare, which differ from the more typical focus on e-commerce \cite{banskotaRecommendingVideoGames2020,mauroPersonalizedRecommendationPoIs2020,peraAutomatingReadersAdvisory2014,tranRecommenderSystemsHealthcare2021,gomezWinnerTakesIt2021}. Similarly, platforms like YouTube\footnote{\url{https://www.youtube.com/myfamily/}} and Spotify\footnote{\url{https://support.spotify.com/uk/article/spotify-kids/}} have customized their recommendations to better serve users like children \cite{papadamouDisturbedYouTubeKids2020}. In these cases, evaluation strategies have also been adapted, incorporating assessment criteria that best fit these unique users and contexts \cite{gomez2021evaluating}. However, a common denominator in these efforts has been the lack of direct involvement from stakeholders who can provide an ``insider perspective,'' as they are the ones who are best positioned to express their needs or who have roles and perspectives specific to the particular recommendation context. 

\subsection{Social Goods and Harms}

In addition to effectiveness or utility for a range of stakeholders, designers have long considered the various types of social impacts --- both good and harmful --- that recommender systems may have.

Some social benefits are closely aligned with the historical core goal of recommender systems to match users with products, information, artistic works, or other creations.
For example, a system that effectively matches music listeners with artists whose work matches their nuanced tastes may provide exposure to a broader set of similar artists than a system that over-emphasizes popularity or demographics \citep{ungruhPuttingPopularityBias2024, flederBlockbusterCulturesNext2009}.
These types of benefits must still be measured and deliberately pursued, as the system may appear to be performing well on some metrics while failing to provide the benefits desired by various stakeholders \citep{ekstrandDistributionallyinformedRecommenderSystem2024}.
This process is made easier because it is an extension, rather than a realignment, of the objectives of recommender systems, i.e., these benefits are in line with recommendation effectiveness as long as the effectiveness is equitably distributed.
To date, several lines of work have explored fairness \citep{fnt-fairness, deldjooFairnessRecommenderSystems2023, wangSurveyFairnessRecommender2023, zehlikeFairnessRankingPart2022}, marketplace equity \citep{mehrotraFairMarketplaceCounterfactual2018}, information access equity \citep{venkatasubramanianFairnessNetworksSocial2021}, and other social impact and social welfare dimensions of recommendation.

In other work, social impacts are studied as additional or even attenuating objectives on recommendation, ensuring that a system advances social objectives while meeting the needs of users and other stakeholders.
Examples include research on harm-aware recommendation \citep{tommaselSecondWorkshopOnline2021} and on ensuring that a wide range of users can use the system \citep{peraAutomatingReadersAdvisory2014,miltonHereThereEverywhere2019,ngRecommendingSocialinteractiveGames2018,mauroSupportingPeopleAutism2022,schedlOnlineMusicListening2019,santos2014extending}.

Another type of social recommender work directly includes beneficial social impacts as general objectives to guide design and objective functions for model optimization and evaluation.
Recommender systems in this category are designed to promote social objectives such as energy savings \citep{starkeEffectiveUserInterface2017}, sustainability \citep{tomkinsSustainabilityScaleBridging2018, patroSafetySustainabilityDesigning2020}, elder safety and quality of care \cite{gutierrezExplainingCallRecommendations2022}, and access to capital \citep{Burke2020AlgoFairness,smithManyFacesFairness2023,burke2022performance}.

While there has been important progress, significant work remains to ensure that recommender systems promote healthy, sustainable societies.
Directly related to our present argument, there is also an urgent need to consider who is involved in identifying social objectives and designing and evaluating recommender systems against them.
\section{Participatory Design and Co-Design}
\label{sec:codesign}

Participatory design (PD) is both a \emph{methodological approach} and a \emph{political stance} that centers on shared decision making and the redistribution of design power. Originating in Scandinavian labor movements to address workplace inequities, PD was designed to incorporate marginalized voices in the design process. This section introduces PD and co-design, including their roots, their application to systems that incorporate significant artificial intelligence (AI) or machine learning (ML) components, and the need for recommender designers to not only adopt PD as a method but also to mindfully consider the politics of the design and design process.

\subsection{History and Principles of Participatory Design and Co-Design}

PD can be best described as a set of analytic and constructive commitments that are either political or methodological, depending on the PD version being used~\cite{randall2007fieldwork}. Because PD began as part of the Scandinavian workplace democracy movement, the two foundational commitments have always been (1) a broad concern with the politics of design and (2) user participation. Early Scandinavian PD projects focused on supporting and strengthening trade unions to provide workers with an opportunity to influence their work situation and use of technology in the workplace~\cite{gregory_scandinavian_2003,kensingParticipatoryDesignIssues1998,bjerknesUserParticipationDemocracy1995}. Later projects were also based on the values and ideas from the trade union projects, but focused on the use of computers within organizational contexts. Over time, the Scandinavian system development tradition has adapted to the shifts in organizational development (e.g., from the employee to the customer), the changing role of unions, and new technological advances, while still contributing to democracy in evolving workplaces~\cite{bjerknesUserParticipationDemocracy1995}.

In human-computer interaction (HCI) research, PD has been described as a method for building “third spaces,” or “hybrid spaces,” that are neutral and unfamiliar to both participants and researchers~\cite{mullerParticipatoryDesign1993,muller2012participatory}. The most common setting for PD research is the participatory workshop. The workshop forms and facilitates a hybrid space by allowing participants and researchers to communicate and reach mutual understanding. To engage users during the workshops, researchers can use a range of PD techniques, methods, and practices developed over the years, including ethnographic methods, storytelling, semi-structured conferences, low-tech prototyping, mock-ups, video and storyboard prototyping, and envisioning future solutions~\cite{muller1993taxonomy}. These tools and techniques can help mediate interactions among the participants, building on their specialized knowledge~\cite{brandt2012tools,ehn2017scandinavian}. By offering participants both agency and opportunities to contribute to system design, PD approaches ensure that users hold equal status with designers throughout a democratic design process~\cite{kensingParticipatoryDesignIssues1998}. The HCI community has adopted PD for its ability to engage diverse stakeholders --- especially individuals from marginalized and underrepresented groups~\cite{haimsonDesigningTransTechnology2020,yipExaminingAdultchildInteractions2017} --- in shaping future design agendas and influencing the development of technologies that will have a direct and immediate impact on people's lives. To date, HCI researchers have used the PD approaches for designing systems and user experiences in various application domains, including e-government and public organization~\cite{pilemalmParticipatoryDesignEmerging2018}, immigration~\cite{duarteParticipatoryDesignParticipatory2018}, healthcare \citep{millerPartnersCareDesign2016, kusunokiSketchingAwarenessParticipatory2015} and emergency medical services \citep{kristensenParticipatoryDesignEmergency2006}, and smart cities~\cite{goochAmplifyingQuietVoices2018}.

\begin{figure*}[tb]
    \centering
    \includegraphics[width=\textwidth]{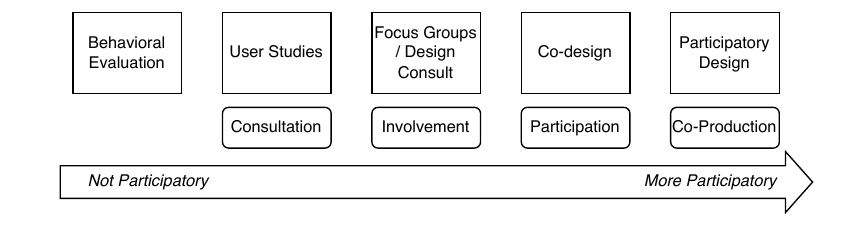}
    \caption{Spectrum of the level of participation in various research and design paradigms.}
    \Description{
        A block diagram of participation, with an arrow underneath going from Not Particpatory on the left to More Participatory on the right.
        The first block on the left is Behavior Evaluation, followed by User Studies (labeled Consultation), Focus Groups / Design Consult (Involvement), Co-design (Participation), and Participatory Design (co-production).
    }
    \label{fig:spectrum}
\end{figure*}

In its early stages and through the mid-1990s, the concept of power was central to PD. Power was defined as a mechanism that regulates decision-making during the design process~\cite{bratteteigDisentanglingPowerDecisionmaking2012}. Making decisions was essentially seen as exercising power. In the traditional PD process, power is grounded in different resources and distributed across stakeholders. For example, a project leader (i.e., a participatory design researcher) with a strong, long-term research vision and with knowledge of users and their work practices, may initiate decisions about the values and concepts, the composition of the research team, and resources allocations. The designer team is responsible for the technical design decisions, while domain experts make decisions about domain-specific issues. The domain experts also negotiate about design implementations with the outside world using their influence and connections. Finally, lay users (i.e., workshop participants) exercise their power through interacting with the prototypes, influencing the researcher's vision and designer's views of what they had designed~\cite{bratteteigDisentanglingPowerDecisionmaking2012}.

The question of power is often overlooked in contemporary discussions of PD ~\cite{bannonReimaginingParticipatoryDesign2018}. The concept of ``user empowerment'' that calls for people to be more in control when using technology ~\cite{johannsenEmpowermentReconsidered2005} is distinct from user empowerment as a design framework because it mostly concerns the power of using technologies that are already designed and deployed. In the context of user empowerment, \citet{sebergerEmpoweringResignationTheres2021} distinguish between ``power to'' and ``power over.'' The ``power to'' concept has been defined as an individual’s capacity to act ~\cite{arendt2019human}. The power to do something within a specific context does not necessarily change the context in which the power is granted or achieved. In contrast, the concept of ``power over'' is based on a more traditional sociological understanding of power as the ability to compel others to act according to one's wishes. The power over reflects inherent power imbalances and constrains the context within which ``power to'' can be exercised by individuals ~\cite{sebergerEmpoweringResignationTheres2021}.
Participatory design seeks to share ``power over'' not just the technology itself, but the process through which it is designed and evaluated.

The political implications of balancing power between stakeholders throughout the PD process are complicated.
Several approaches to balancing power structures in a participatory design process have been proposed. One approach is to collaboratively develop power-balanced processes with the community, fostering trust between designers and social actors and allowing the community to guide decisions and share power ~\cite{tomasinigianniniPowerbalancedParticipatoryDesign2022}. This approach necessitates a reinterpretation of the designer's role, promoting the de-learning of traditional practices and fostering dialogue. A power-balanced PD process views the community as an equal partner rather than merely an end-user. Another approach encourages reflection and action, allowing periodic evaluations of power dynamics to prevent biases and oppressive attitudes in the collaborative effort~\cite{freirePedagogyOppressed1978}. As a third option, a power relation triangle has also been proposed, which includes participants, designers, and external conditions, with decision-making positioned at the core of the interactions for understanding power imbalances ~\cite{volkmannBalancingPowerRelations2023}. In this approach, external factors may impact power relations and shifts in initiative can alter the power dynamics.

\emph{Co-design} is a related participatory approach to designing technologies and systems. Although participatory design and co-design share many commonalities and are often used interchangeably, they represent two different approaches to designing systems based on the level and nature of user participation.  The participatory power of users in design research can be seen as a spectrum, starting with \textit{consultation}, where users are informally asked for opinions or preferences, to \textit{involvement} or more substantive opportunities for users to contribute to decision making, to \textit{participation} or allowing users to guide the design process and even define the agenda, to \textit{co-production}, where participants have an equal say in deciding the project goals and outcomes ~\cite{delgadoParticipatoryTurnAI2023,sandersCocreationNewLandscapes2008} (Figure ~\ref{fig:spectrum}). User participation in co-design is typically at the \textit{participation} level, whereas user participation in PD usually reaches the \textit{co-production} level because PD also involves the political component within the larger socio-technical system being designed.

\subsection{Co-Designing AI-Based Systems}
\label{sec:codesign-ai}

In recent years, researchers have explored the application of PD in designing AI-based technologies as ``participatory AI''~\cite{delgadoParticipatoryTurnAI2023}. Because emerging technologies such as AI can negatively affect people's lives through algorithmic bias, opaque decision making, and the spread of misinformation~\cite{demartiniHumanintheloopArtificialIntelligence2020}, applying participatory approaches to designing AI-based systems is critical.

``Participatory ML''~\cite{kulynych2020participatory} has emerged as a co-design approach for considering the main components of ML systems from a human-centered perspective. Critical considerations across different stages of ML design and development include whether (1) the datasets are ecologically valid for the targeted population; (2) the algorithms are grounded in human theory, understanding, and knowledge; (3) the evaluation metrics are including both quantitative and qualitative metrics; and (4) the artifacts and outcomes are safely deployed in real-world settings based on human values~\cite{raziHumancenteredSystematicLiterature2021}.  
PD approaches have been used to address each of these critical aspects of ML design, including dataset building~\cite{pushkarnaDataCardsPurposeful2022}, design and validation of ML algorithms~\cite{theodorouDisabilityfirstDatasetCreation2021}, and understanding of algorithmic accountability~\cite{queerinaiQueerAICase2023} and equity~\cite{katellSituatedInterventionsAlgorithmic2020}.

Adopting PD for designing AI-based systems across different application domains raises several methodological questions ~\cite{zytkoParticipatoryDesignAI2022}:

\begin{enumerate}
    \item Who will conduct participatory AI design research?
\item Which aspects of AI can or should be co-designed?
\item What will be the process of user participation in AI co-design?
\item What artifacts will be used in the process?
\item How will the collaboration between co-designers and researchers with different domain expertise be facilitated?
\item What are the persistent challenges to participatory AI design?
\end{enumerate}

Recent AI co-design efforts have been criticized for minimal engagement activities with users through a single design workshop or focus group aimed at obtaining feedback on designer-generated artifacts~\cite{zytkoParticipatoryDesignAI2022}. This superficial level of engagement, only reaching the involvement level of participation (Figure~\ref{fig:spectrum}), diminishes the fundamental objective of PD: to foster active and sustained involvement of user and influence the final system design through a highly interactive and iterative process~\cite{zytkoParticipatoryDesignAI2022}.
Ongoing engagement and sharing of power is also a key distinction between fully participatory AI and human-centered AI, as articulated by \citep{shneidermanHumancenteredAI2022}, which focuses on consulting with users through traditional HCI techniques rather than fully involving them as co-producers of AI.
Other critiques of practices in participatory ML include equitable work with participants~\cite{corbettPowerPublicParticipation2023,delgadoParticipatoryTurnAI2023,fefferPreferenceElicitationParticipatory2023}, ``participation-washing'' \cite{sloaneParticipationNotDesign2022}, and co-option of participatory work~\cite{birhanePowerPeopleOpportunities2022}.

A recent interview study of ``participation brokers'' who lead participatory ML projects highlighted the need for equitably balancing the value generated through participatory ML with educating users and advocating for participant rights~\cite{cooperFittingParticipationForging2024}. Alleviating frustration from indirect stakeholders by converting qualitative stories into main ML processes is also important in the participatory context~\cite{cooperFittingParticipationForging2024}. Several efforts have been evaluated that use generative AI to replace human participation, thereby reducing the cost of research and development and increasing the diversity of collected data~\cite{dillionCanAILanguage2023}. However, replacing research participants with AI-generated data conflicts with the essential values of representation, inclusion, and understanding of the human~\cite{agnewIllusionArtificialInclusion2024,hardingAILanguageModels2024}.

PD and co-design offer more than methods for creating systems --- they represent a commitment to equitable design practices that center diverse voices and share decision-making power. While AI/ML-intensive systems present unique challenges, integrating participatory approaches requires not only methodological shifts but also a fundamental rethinking of design politics. Adopting such commitments in recommender system design can lead to more inclusive, accountable, and impactful innovations that serve diverse user bases and advance democratically-developed social good.
\section{Participatory Recommendation}
\label{sec:corecsys}

Our central argument (outlined in \S\ref{sec:intro}) is that recommender systems --- particularly those aimed at social good --- should be designed \textit{with}, not just \textit{for}, their users, providers, society at large, and various specific people and groups impacted by the system.
Under this approach of \textit{participatory recommendation}, designers and other stakeholders work together in a \textit{participatory design team} (PD team) to design and evaluate a recommender system.
Participatory methods are important for all types of recommendation goals, not just those aimed at social benefit. However, the explicit role of values in determining and pursuing social good heightens the necessity of participatory methods because they provide a means for ensuring the values incorporated into the system come from the community in which it will operate.

A commitment to participatory recommendation will pervade the entire recommender system design and evaluation process.
The objectives for the recommender system, the computational and interactional mechanisms used to meet those objectives, and the evaluation of whether the system achieves its objectives should all be developed in collaboration with the system's beneficiaries.
This participatory collaboration is necessary not only for the kinds of individual benefit or utility objectives frequently considered in recommender system design but also for objectives related to social impact and welfare.
PD team collaborations should be genuinely power-sharing relationships in which the people who will use and be impacted by the system are not only consultants or subjects in user studies, but have meaningful control over the final product and its initial and ongoing evaluation.
Practically, participatory recommendation extends the arguments by \citet{ekstrandBehaviorismNotEnough2016} to avoid purely behaviorist approaches to designing and evaluating recommender systems, providing a realization of the argument made by \citet{burke:post-userist} that the recommender systems community must attend to the relationships between various parties in a recommendation interaction.

Human-centered and user-centered approaches to AI generally \citep{shneidermanHumancenteredAI2022} and recommender systems specifically \citep{konstanHumanCenteredRecommenderSystems2021}, grounded in traditional human-computer interaction methods of user-centered design and evaluation through user studies, provide a valuable starting point for participatory recommendation.
However, realizing a participatory vision of full co-production of recommender systems, or even participatory co-design, requires more.
First, more stakeholders should be involved. In addition to including users, participatory recommendation design needs to involve producers and other people affected by the recommender system's operation.
Second, traditional human-centered methods retain power, particularly design and evaluation authority, with the designers, whereas participatory recommendation shares that power between the designers and other stakeholders. Finally, while ``control'' is important in both visions, human-centered AI is primarily concerned with designing systems that allow user control over their operation, while participatory recommendation and participatory AI envision ensuring users and other stakeholders have power in the design and evaluation process itself.

Participatory recommendation is not an \emph{easy} agenda to pursue (technically, organizationally, or politically), but it is necessary to truly achieve the goals of recommendation for social good, and in turn, recommendation for general human benefit.
Ursula Franklin [\citeyear{franklinRealWorldTechnology2004}] distinguishes between \textit{relational} technologies built on reciprocity and mutual respect and unidirectional \textit{broadcast} technologies.
Too often, recommender systems operate as broadcast technologies, but we posit that participatory approaches are key to unlocking recommender systems as a relational technology.
Relational understandings of recommendation are not new, as they were articulated in the field's early days by \citet{hillRecommendingEvaluatingChoices1995} and more recently highlighted by \citet{burke:post-userist}.
In an era of techlash \citep{atkinsonPolicymakersGuideTechlash2019}, we also hypothesize that people will be more accepting and trusting of recommender systems if they or their communities had a meaningful say in creating them \citep{ekstrandBehaviorismNotEnough2016}.

Participatory recommendation will likely need to start with small-scale systems, aimed at solving information access and discovery problems for small, well-defined use cases.
Lessons from designing smaller systems will inform the methods needed to design or re-design global systems. Global designs will likely arise as amalgamations of approaches from many smaller, more localized design efforts because of the challenges in representing a global population of users, sellers, subjects, or others in a manageable participatory design setting.
A robust marketplace of small- to medium-scale information ecosystems and e-commerce platforms may also be socially beneficial for other reasons \citep{rajendra-nicoluccisThreeleggedStoolManifesto2023}.

Effective design, development, and evaluation will iterate between the various stages, in close collaboration with people affected in different ways and with different degrees of power.
This collaboration is not unidirectional, but a true collaborative effort with shared power.
The recommender system designers should also not be reduced to stenographers of designs produced by others, as they bring significant expertise on technical possibilities and the computational and systemic consequences of potential designs.
In a participatory approach, designers are heavily involved but do not serve as the final arbiters of the design. Instead, they contribute their expertise to a conversation and collaboration with people who will use and be impacted by the design.
Implementing participatory recommendation requires careful consideration of the power of different people involved in the processes. Access to data and computational resources are also a dimension along which designers and the platform employing them may have crucial power not held by other participants in the design process.

We next describe how different elements of the recommender systems development lifecycle can be reenvisioned as a participatory process.
Our discussion here is forward-looking, as much of the work on adapting and applying participatory design and co-design, and developing the technical and methodological capabilities to support these methods in information access and AI systems has yet to be done.

\subsection{Conception and Problem Definition}
\label{sec:corecsys-definition}

Recommender system design begins with conceiving the initial application and defining the problem that the system would address. 
A participatory approach will start by defining the problem with people who will use or be affected by the system, without pre-supposing that a recommender system is necessary or appropriate.
This approach applies to greenfield platforms or applications and to new or enhanced recommendation capabilities in existing systems.

Participatory recommendation starts from a place of humility and begins with a page from information science, information retrieval, and HCI: understanding what users and other stakeholders \textit{need}.
Music recommender systems can serve as a simple example.
Users may need to know when their favorite artists have released new music.
Depending on the platform dynamics, artists may desire to have their newer music surfaced if current listening and recommendation patterns favor their older hits.
The platform operator needs their business to generate enough profit to be sustainable.
Developers may have a range of needs, from continued employment to an intrinsic interest in supporting music discovery.
Although all of these needs could point to a recommendation experience that helps locate new music, participatory methods will help elicit these needs and resulting designs directly from the platform stakeholders.

Some of the participatory engagement at this stage of the recommendation design can be done at the \textit{consultation} level of participation (Figure~\ref{fig:spectrum}), such as working with focus groups of users or artists.
Participatory methods begin to differentiate from this consultation as more people are involved in the later stages, and as the recommender system designers commit to collaborate with the community participants.

Another key part of defining the problem is defining the stakeholder groups and recruiting participants for the design process.
Stakeholder identification will often start with the designers' ideas of what perspectives are important for a particular recommender system. Designers should engage in that planning, as the goals of participatory collaboration with shared power suggest inviting the early participants to recommend additional people or groups that are not yet accounted for.
In such designer-originated participatory recommendation projects, perhaps designed to make existing human-centered research and development more participatory, participant recruitment and compensation will most likely happen through existing organizational channels or through building such mechanisms where they do not yet exist.
In some cases, however, the recommender system concept may originate from a stakeholder group themselves, such as an artist collective wanting a recommender system to help clients or patrons locate and select from the work of the different artists.
Such ground-up, community-originated efforts, where they exist, are a clear realization of the full co-production level of participatory design, and already come with at least one set of stakeholders and participants.

In both problem definition and the remaining phases of participatory recommender systems research, different participants may have conflicting goals or design suggestions. They may also propose ideas that expand beyond the capability of current systems or with negative social or system consequences.
Navigating these conflicts through collaboration is vital to ensuring a truly participatory design and evaluation process, although the precise details will depend on the context and application.
\citet{rauWhyInteractiveLearning2013} provide one approach to resolving conflicting design goals through identifying hierarchies of goals that may be useful in participatory recommendation, as long as that hierarchy itself is an outcome of the participatory collaboration rather than imposed by the designers.

\subsection{Design}
\label{sec:corecsys-design}

Participatory design and co-design make a clearer difference as we move from problem definition to the actual design of the recommender system and its surrounding application and user experience.
Once the PD team has collaboratively agreed on the need for a particular recommender system, the actual design proceeds in continued collaboration.

Co-designing a recommender system will involve meaningful input and decision-making authority from multiple stakeholders on all aspects of the design, including:

\begin{itemize}
    \item Location of the recommendations within the user workflow.
    \item Criteria for item inclusion and prioritization.
    \item User interface elements (display, layout, feedback or control mechanisms).
    \item Data used (or not used) as a basis for recommendation.
    \item Permission structures around opting in or out of the recommendations (for both users and providers).
    \item Key principles of evaluation and objective functions.
\end{itemize}

We discuss the last point above --- evaluation and objectives --- in more detail in the next section, but as recommender model behavior is driven in large part by the designs of its objective functions and the evaluations by which it is assessed and optimized, these metrics and analyses should be collaboratively developed by the PD team.
\citet{schellingerhoutCodesignStudyMultistakeholder2023} have made some early progress in co-designing aspects of recommender systems, finding that different stakeholders in a system may have very different needs from its outputs (in their study, explanations).

The degree of agency and ability the different members of the PD team have to shape the final outcomes determines where on the participatory spectrum a particular recommender system design effort will fall (Figure~\ref{fig:spectrum}).
Co-design focuses on collaboration during the design phase, involving stakeholders (e.g., users or communities) to contribute their perspectives and ideas and provide feedback, reaching the ``participation'' level. Full co-production includes multiple stakeholders as equal partners across all stages of a project.
Both of these levels represent an improvement in participation over standard recommender system design practice.

\subsection{Evaluation}
\label{sec:corecsys-eval}

Evaluation is a key aspect of recommendation, particularly its algorithmic and modeling components.
As noted in \S\ref{sec:corecsys-design}, it is deeply intertwined with other design efforts: designing an evaluation and related objective functions (for model training, hyperparameter optimization, design evaluation, etc.) is a part of recommender system design.

Substantial research is needed to determine how to effectively co-design --- and ``co-evaluate'' --- recommender systems.
Metrics and evaluation setups are complex, mathematically nuanced, and statistical (and sometimes directly stochastic).
Therefore, they can be difficult even for experienced recommender system designers to interpret.
Participatory recommendation during the evaluation phases will robustly involve people with various perspectives on the system and widely varying levels of expertise in other topics, but rarely in recommender systems.
Facilitating this involvement will likely require advances in both participatory protocols and computational evaluation methods.
The full PD team can and should be involved in many different ways, including:

\begin{itemize}
    \item Qualitatively evaluating the system based on their own experience.
    \item Designing and validating evaluation metrics.
    \item Reviewing and selecting evaluation data and data preparation steps (e.g., integration, splitting, setting up simulated tasks).
    \item Designing online evaluation (e.g., what metrics in an online field trial would indicate system success).
    \item Reviewing evaluation results to assess whether the system meets the problem and design parameters (as discussed in \S\ref{sec:corecsys-definition} and \S\ref{sec:corecsys-design}, respectively).
\end{itemize}

Some aspects of the system can be directly and qualitatively assessed by design participants, particularly user interface and directly user-observable aspects (e.g., inclusion or exclusion of appropriate items, or perceived diversity). Many teams already do this internally by reviewing their own recommendations. Expanding this initial user involvement to a broader set of representatives will improve the evaluation.
Other aspects of recommender system behavior and impact are harder to directly observe, either because their effects are behind the scenes or they are statistical (and therefore cannot be assessed from individual recommendation results or interactions).
Fairness objectives are one class that often fall into this category: looking at a single recommendation list, or even many recommendations for a single user, is not sufficient to assess many reasonable fairness criteria, such as equal opportunity for content providers \citep{diazEvaluatingStochasticRankings2020,biegaEquityAttentionAmortizing2018}.

Fully participatory co-production of a recommender system arises when the user and other stakeholder authority is equal to that of the recommender system designers in deciding whether the evaluation shows the system is fit for purpose and should be deployed (or not).
\citet{smithRecommendMeDesigning2024} incorporated recommender system practitioners in the process of co-designing recommender system evaluation, but much more work is needed to incorporate additional stakeholder participants and account for power differentials and stakeholder conflict.

\subsection{Monitoring and Review}
\label{sec:corecsys-review}

Recommender system design and evaluation is not a one-and-done process, but rather an iterative one in which the system is continually reviewed and improved, and is monitored for changes in effectiveness or behavior as new content, users, interactions, and other data arrive.

Building on the iterative nature of participatory design, participatory recommendation should incorporate participatory input and ongoing co-design efforts in at least three ways:

\begin{enumerate}
    \item Designing the ongoing monitoring and evaluation procedures to be carried out by the platform. What needs to be measured and monitored to ensure the system's ongoing fidelity to the agreed-upon problem definition and design?

    \item Regular review of system behavior and emerging impacts, not just by the designers but by users.

    \item Conception and design of improvements, enhancements, or removals of the design based on ongoing experience and community input (e.g., what users hear from other members of their communities).
\end{enumerate}

Ongoing review and revision by the full PD team will ensure that the recommender system continues to meet the needs of the many different people with an interest in its operation and behavior.
\section{Participatory Approaches in Practice}
\label{sec:cases}

To demonstrate our argument, we discuss how participatory methods could impact the design and evaluation of recommender systems in several application contexts.

\subsection{Fair Multi-sided Marketplaces}

Many platforms employing recommender systems are some form of \textit{multi-sided marketplace}, connecting people to each other through the products, media, or opportunities being recommended.
In these settings, people have different interests and needs with regard to the recommender system's usefulness, fairness, and other properties, depending on their role in any particular transaction or interaction \citep{sonboliMultisidedComplexityFairness2022}.
A recent line of work has proposed methods for human-centered operationalization of recommendations. This approach acknowledges that fairness is a contested concept that will have different meanings to different individuals and begins with exploratory semi-structured interviews eliciting how fairness issues arise in everyday experiences of users. One example is an interview study with employees of the crowd-funded microlending platform Kiva Microloans \citep{smithManyFacesFairness2023}. The findings showed a wide range of fairness logics articulated by the employees, who, in their organizational roles, had greater or lesser interaction with different stakeholders served by Kiva: lenders, borrowers, partner organizations, corporate sponsors, and others. Although using a human-centered approach, this study did not incorporate actual end users (recommender consumers) of Kiva, whose concepts of fairness were examined in a previous study \citep{sonboliFairnessTransparencyRecommendation2021}. The same approach was pursued in an interview study with journalists that looked into the way journalistic values are (or are not) being realized in news recommender systems  \citep{alaqabawyItsJustRobot2023}. 

While interviews and subsequent analysis help recommender system designers develop general categories and logics of fairness, further steps are needed to fully operationalize and implement the ideas identified through interviews.
It is up to system designers to articulate the concepts users express in computational terms. 
This process of operationalization is uncertain because crucial nuances, obvious but implicit for practitioners, might be lost in the (technical) translation. 
Once system designers have developed mathematical definitions of proposed fairness metrics, these can be brought back to stakeholders for discussion in focus groups.
\citet{smithRecommendMeDesigning2024} illustrate this two-step focus group process in the context of provider-oriented fairness for recommendations in social media and dating sites. In addition to examining designers' fairness measures, \citeauthor{smithRecommendMeDesigning2024} encouraged providers to formalize their own fairness notions.

In participatory recommendation, participation in the assessment of fairness continues through the design and implementation of transparency mechanisms. Although considerable research exists on recommendation explanation, it is heavily focused on the explanation of individual recommendation results \cite{tintarevExplainingRecommendationsDesign2015} and not on the types of insights into overall system behavior required by a participatory design effort. 
In addition, explanation and transparency for stakeholders other than recommendation consumers, such as providers of recommended items, are rarely studied in recommender research. \citet{varastehComparativeExplanationsRecommendation2024} discuss some of the challenges that would be involved in developing provider-side transparency mechanisms. Still, such tools would be essential for ongoing monitoring by providers throughout a recommender system's lifecycle. 
Research on operationalizing fairness objectives in collaboration with non-developer stakeholders, or with multiple stakeholder groups in a single effort, is also limited.

\subsection{Social Media and Online Safety}

Recommender systems are influential in shaping user experiences in \textit{social media} and influencing what content users see and interact with. However, the mechanisms that make these systems engaging and personalized also raise critical concerns about their impact on user \textbf{online safety} and their potential to amplify harmful content.
Recommender ranking algorithms, often prioritizing engagement over safety, can inadvertently expose users to harmful, inappropriate, or distressing content. This exposure can range from explicit content like violence and hate speech to subtler forms of harm like misinformation, stereotypes, and content that promotes unhealthy behaviors~\cite{ungruhAhThatsGreat2024}. 
For more vulnerable users such as children, the exposure to harmful content can lead to ``harmful pathways,'' where users encounter increasingly extreme content over time~\cite{ungruhAhThatsGreat2024}.
For instance, for individuals in recovery from conditions like eating disorders, encountering pro-eating disorder content through recommendations can be particularly distressing and can even lead to relapse. The challenge is that these types of recommender systems often struggle to discern nuanced contexts and may associate pro-eating disorder content with recovery content, leading to harmful recommendations~\cite{golbeckRecommenderSysteminducedEating2025}.

As in other domains, social media recommender systems can exacerbate existing social and economic inequalities, exposure to misinformation, hate speech, and harmful stereotypes.
The opacity of many systems, particularly those using complex AI models, makes it difficult for users to understand how these systems work and why they are being recommended certain content.
This lack of transparency limits users' ability to control their online experiences, challenge potentially biased or harmful recommendations, and hold platforms accountable for the outcomes of their algorithms~\cite{smithRecommenderSystemsAlgorithmic2022}.

Participatory approaches will involve a wider range of stakeholders to collaboratively understand and address the potential harms of recommender systems in social media.
Participatory design can be used to develop explanations for why certain content is being recommended and can empower users to make informed choices and identify potential biases or harms.
Using co-design strategies, users could ensure more control over their recommendation settings, allowing them to customize content preferences, filter out specific topics, or opt out of certain types of recommendations~\cite{smithRecommenderSystemsAlgorithmic2022}.
Participatory and co-design approaches involving various stakeholders can help identify and define potential harms that may not be apparent to developers or researchers. Stakeholders can provide valuable input on developing metrics that accurately reflect the values they care about. In addition, the lived experiences and insights of users can inform the development of interventions and mitigation strategies that are both practical and responsive to their needs and the needs of their communities.

\subsection{Education}

The ideas of participatory recommendation are not limited to recommender systems, but are applicable to a range of information access technologies, including search engines and natural language tools for information tasks.
Several of the authors have been working on developing recommender systems and search tools for \textit{educational settings}, particularly to support K--12 teachers in finding reading materials for their students~\citep{literate,peraHowCanWe2014,jasonhallUBRBookSearch2017,murgiaSevenLayersComplexity2019}.

One specific example has as an educational goal supporting children's literacy development and proficiency by engaging them with diverse \textit{informational texts} in their early education \citep{literate}.
The majority of texts read by adults are informational \citep{atkinsonMakingScienceTrade2009}, and comprehension of these non-narrative, nonfiction texts is key to academic success \citep{dukeReadingWritingInformational2003}. The degree of text ``authenticity'' --- whether the author wrote the text purely for pedagogical purposes (e.g, a textbook) or for a wider, real-world audience (e.g., a newspaper) --- also significantly impacts learning. For example, \citet{purcell-gatesLearningReadWrite2007} found that elementary children who engaged in authentic science literacy tasks were better able to comprehend and produce informational texts in later grades.

While research provides guidance on best practices on using informational texts in the classroom \citep[e.g.,][]{malochExposureUsesInformational2008, bradleyInformationBookReadAlouds2010}, it fails to address the challenge of finding these readings. 
\citet{literate} found that existing search technology does not suffice when teachers are looking for authentic informational texts for classroom use.
In the reported interviews, teachers frequently discussed that they could not find what they were looking for on Google, and that typical search engines often led them to materials blocked by a paywall.
This work also highlighted several specific problems that can serve as a starting point for technical innovation --- current search and recommender systems are not designed to support teachers in identifying resources that are relevant, appropriate, and accessible to their students. In addition, the results showed teacher interest in a recommender system that would (1) allow them to locate a variety of texts that match student interests, (2) filter for readability, and (3) support their ability to locate free resources that match curriculum needs.

Fully realizing this kind of information access system in a manner that \textit{empowers} teachers and integrates with their work and teaching contexts demands a participatory approach in which teachers are full collaborators --- if not leaders --- in its design and evaluation.
Because teachers are experts in both the curriculum and pedagogy of their classrooms, they are best positioned to accurately identify what will best serve their instructional needs, both in terms of the recommended texts and the technology.
The educational technology market is replete with technologies that have been developed and ``thrown over the wall,'' designed to serve the interests of administrators and technologists instead of students.
Many are built on surveillance-capitalist technology paradigms \cite{hillmanHowSurveillanceCapitalism2022, stockmanSurveillanceCapitalismSchools2022}, bypassing teachers by connecting directly with students (and harvesting data from them).
Re-envisioning the information access and educational technology design process around participatory design with teachers at the center is a promising approach to delivering technology that will integrate well with classroom activities and meet real, on-the-ground needs faced by teachers and students.
It will also result in technology that is better adapted to the particularities of different teaching contexts, easing adoption and improving impact.
Teachers' involvement in design should not be limited to visuals, layouts, and high-level features, but also to designing the machine learning objectives and, ultimately, the metrics and experimental procedures that will be used to design and iterate on the technical advances needed to bring teachers’ design ideas to life.

\section{The Vision of Participatory Recommendation}
\label{sec:vision}

We envision a future where people are meaningfully involved in designing and evaluating the recommender systems that affect them.
The participatory agenda we have described is not limited to recommender systems, but also applies to other information access systems, such as search engines and informational AI systems.

The vision of participatory recommendation includes two major components:

\begin{itemize}
    \item  \textbf{Methodological changes} to incorporate participatory methods into recommender system design, evaluation, and deployment.
    \item \textbf{Political commitments} about how power is allocated, whose opinions on recommender systems matter, and who is empowered to meaningfully affect both individual systems and the broader landscape of recommendation, retrieval, and personalization.
\end{itemize}

These methodological and political moves are particularly important in the pursuit of recommender systems for social good because such efforts necessarily entail judgments about what constitutes a ``social good.'' To build recommender systems that truly empower communities and societies to pursue good, it is necessary to use meaningful participatory approaches in which designers relinquish power in determining what is ultimately built.
Traditional design paradigms risk building systems that paternalistically impose visions of social good defined by organizations (or individuals) that do not share the values of those most directly impacted by the system or of well-meaning systems harming the communities they are intended to support.

We do not diminish or devalue technical work aimed at social goals. Directly pursuing social goals through technical research and development, in addition to establishing and evaluating technical prerequisites that can be applied to positive social outcomes, are valuable and necessary.
We hope to see such work grounded and contextualized in a robust understanding of social good, including how that notion of social good was derived. We also hope to see any claims of actual attainment of social good assessed by the people who will be affected --- positively or negatively --- by the proposed good.

Embracing the challenge of participatory recommendation requires extensive research on several important points, including those listed in \S\ref{sec:codesign-ai} for participatory AI in general.
Key research needs include:

\begin{itemize}
    \item Tools and methods, validated across multiple applications, to support participatory design and co-design of recommender systems, information retrieval, and similar technologies.
    Existing methods are highly applicable and recent efforts such as participatory AI playbooks \citep{pushkarnaDataCardsPurposeful2022, changExpansiveParticipatoryAI2022} will be useful for recommender systems; new or revised tools may be necessary to support effective co-design.

    \item Evaluation techniques, metrics, and reporting concepts, along with participatory protocols using them, that enable non-experts to meaningfully assess a recommender system's performance or behavior.

    \item No-code or low-code recommender system design tools that allow communities to adjust recommendations to their particular needs.

    \item Adaptable, low-data recommendation techniques that can be deployed and customized without the need to collect extensive interaction data by means typically only available to large platforms.

    \item Accessible, participatory auditing methods and tools to enable people to rigorously interrogate and understand the behavior of the recommender systems they already use.
\end{itemize}

Recommender systems and other information access technologies have significant capability both to advance social good and to catalyze harm.
Deliberate design to effectively meet user needs, provide opportunities for providers, and advance social welfare objectives is vital to ensuring the technology benefits the people it affects.

Although our present argument emphasizes the need, value, and practical guidance for designing recommender systems with a focus on social good, we also recognize that these systems do not operate in isolation and do not serve a single objective. Multiple stakeholders contribute to and influence the requirements of recommender systems, which must ultimately be balanced and prioritized \cite{abdollahpouriMultistakeholderRecommendationSurvey2020,surerMultistakeholderRecommendationProvider2018}. This balancing effort is challenging because there is no one-size-fits-all solution --- the criteria, objectives, and priorities of different participants' concerns vary depending on the specific use case. 

One of the most immediate conflicts in participatory recommendation, or even recommendation for social goals through traditional design and research methods, is the inherent tension between profit-driven incentives and the pursuit of broader social good. Many recommender systems are integrated within commercial platforms that primarily aim to maximize sales, engagement, retention, and ultimately revenue \cite{jannachEscapingMcNamaraFallacy2020}. These financial incentives can sometimes conflict with ethical considerations or with the power-sharing objectives of participatory design. For instance, engagement-driven algorithms may inadvertently prioritize sensationalist or polarizing content because it generates more clicks and user interaction, even though such content can have negative societal implications.
Addressing this tension, along with those arising from multiple stakeholder considerations, is an ongoing discussion. In this paper and our prior work \citep{ekstrandNotJustAlgorithms2024, burkeDecenteringTraditionalUser2025} we are attempting to advance that vital conversation about the importance of balancing diverse needs and perspectives. 

Bringing the participatory turn in AI \citep{delgadoParticipatoryTurnAI2023} to the project of building useful and socially-beneficial recommender systems will give more people the opportunity to directly contribute to these design efforts, and hold such efforts accountable to the people and communities they are intended to benefit.

Information access systems will best serve social good when people who experience their consequences are empowered to participate in their design and evaluation.
We invite you to join us in making participatory recommendation a reality.

\begin{acks}
Robin Burke was supported by the National Science Foundation under grant IIS-2107577.
\end{acks}

\bibliographystyle{ACM-Reference-Format}
\bibliography{zotero}

\end{document}